\documentclass[prl,twocolumn,tightenlines,preprintnumbers,floatfix,nofootinbib]{revtex4}
\usepackage{graphicx,amsfonts,amssymb,amsmath}

\newcommand{\nc}{\newcommand}
\nc{\grad}{\nabla}  
\nc{\tr}{\mathop{\rm Tr}}
\nc{\half}{{1\over 2}}
\nc{\third}{{1\over 3}}
\nc{\be}{\begin{equation}}
\nc{\ee}{\end{equation}}
\nc{\bea}{\begin{eqnarray}}
\nc{\eea}{\end{eqnarray}}

\nc{\dint}[2]{\int\limits_{#1}^{#2}}
\nc{\D}{\displaystyle}
\nc{\PDT}[1]{\frac{\partial #1}{\partial t}}
\nc{\tw}{\tilde{w}}
\nc{\tg}{\tilde{g}}
\nc{\newcaption}[1]{\centerline{\parbox{5.6in}{\caption{#1}}}}
\def\href#1#2{#2} 

\def\beq{\begin{eqnarray}}   
\def\eeq{\end{eqnarray}}

\def\lsim{\mathrel{\rlap{\lower3pt\hbox{\hskip0pt$\sim$}}
    \raise1pt\hbox{$<$}}}         
\def\gsim{\mathrel{\rlap{\lower4pt\hbox{\hskip1pt$\sim$}}
    \raise1pt\hbox{$>$}}}         

\def\Id{\hbox{1\kern-.23em{\rm l}}}

\def\lsim{\mathrel{\rlap{\lower3pt\hbox{\hskip0pt$\sim$}}
    \raise1pt\hbox{$<$}}}         
\def\gsim{\mathrel{\rlap{\lower4pt\hbox{\hskip1pt$\sim$}}
    \raise1pt\hbox{$>$}}}         

\nc{\al}{\alpha}
\nc{\ga}{\gamma}
\nc{\de}{\delta}
\nc{\ep}{\epsilon}
\nc{\ze}{\zeta}
\nc{\et}{\eta}
\renewcommand{\th}{\theta}
\nc{\Th}{\Theta}
\nc{\ka}{\kappa}
\nc{\la}{\lambda}
\nc{\rh}{\rho}
\nc{\si}{\sigma}
\nc{\ta}{\tau}
\nc{\up}{\upsilon}
\nc{\ph}{\phi}
\nc{\ch}{\chi}
\nc{\ps}{\psi}
\nc{\om}{\omega}
\nc{\Ga}{\Gamma}
\nc{\De}{\Delta}
\nc{\La}{\Lambda}
\nc{\Si}{\Sigma}
\nc{\Up}{\Upsilon}
\nc{\Ph}{\Phi}
\nc{\Ps}{\Psi}
\nc{\Om}{\Omega}
\nc{\ptl}{\partial}
\nc{\del}{\nabla}
\nc{\ov}{\overline}
\nc{\gsl}{\!\not}
\nc{\bi}[1]{\bibitem{#1}}
\nc{\fr}[2]{\frac{#1}{#2}}
\nc{\dsl}{\partial\!\!\!\!\!\!\not\,\,}
\nc{\gm}{\mbox{$\gamma_{\mu}$}}
\nc{\gn}{\mbox{$\gamma_{\nu}$}}
\nc{\Le}{\mbox{$\fr{1+\gamma_5}{2}$}}
\nc{\Ri}{\mbox{$\fr{1-\gamma_5}{2}$}}
\nc{\GD}{\mbox{$\tilde{G}$}}
\nc{\gf}{\mbox{$\gamma_{5}$}}
\nc{\Ima}{\mbox{Im}}
\nc{\Rea}{\mbox{Re}}

\nc{\av}{\langle \ph\rangle}
\nc{\ntwo}{${\cal N}\!\!=\!2\;$}
\nc{\none}{${\cal N}\!\!=\!1\;$}
\nc{\nfour}{${\cal N}\!\!=\!4\;$}

\def \bi{\bibitem}
\nc{\rf}[1]{(\ref{#1})}
\def \del{\partial}

\begin{document}
\preprint{hep-ph/0402023}
\preprint{$\;\;\;$DESY 04-020}
\preprint{$\;\;\;$FTPI--MINN--04/02}
\preprint{$\;\;\;$UMN-TH-2229/04}
\preprint{$\;\;\;$UVIC-TH-03-10}
\preprint{$\;\;\;$CERN-PH-TH/2004-017}

\setcounter{page}{1}

\vspace*{0.2in}
\title{ Probing CP Violation with the Deuteron Electric Dipole Moment\\ $\;\;$ \\}

\author{Oleg Lebedev$^{\,a}$, Keith A. Olive$^{\,b}$, Maxim Pospelov$^{\,c}$
  and Adam Ritz$^{\,d}$}
\affiliation{$^a$DESY Theory Group, D-22603 Hamburg, Germany $\;$ \\
$^b$William I. Fine Theoretical Physics Institute, 
University of Minnesota, 116 Church St SE, Minneapolis, MN 55455, USA $\;$ \\
$^c$Department of Physics and Astronomy, University of Victoria, 
Victoria, BC, V8P 1A1 Canada $\;$ \\
$^d$ Theory Division, Department of Physics, CERN,
CH-1211 Geneva 23, Switzerland}

\begin{abstract}
  
We present an analysis of the electric dipole moment (EDM) of 
the deuteron as induced
by CP-violating operators of dimension 4, 5 and 6 including 
$\th_{\rm QCD}$, the EDMs and color EDMs of quarks, four-quark 
interactions and the Weinberg operator. We demonstrate that the precision
goal of the EDM Collaboration's proposal to search for the deuteron EDM, 
$(1-3)\times 10^{-27} e\,{\rm cm}$,  will provide an improvement in sensitivity
to these sources of one-two orders of magnitude relative to
the existing bounds.
We consider in detail the level to which CP-odd phases can be probed within 
the MSSM.

\end{abstract}

\maketitle

\newpage


The most stringent constraints on flavor-diagonal CP violation in 
the hadronic sector arise from bounds on the EDMs of 
the neutron \cite{n}, mercury \cite{Hg}, and in certain cases thallium \cite{Tl}. 
These experiments have important implications for physics beyond 
the Standard Model, and its supersymmetric extensions in particular
(see e.g. \cite{KL}).

In what follows, we will show that a proposed measurement of the deuteron EDM 
\cite{exp}, with  projected sensitivity
\begin{equation}
 |d_D| < (1 - 3) \times 10^{-27}\, e\,{\rm cm},
\label{bound}
\end{equation}
would improve the sensitivity 
to ${\bar \theta}_{\rm QCD}$ and SUSY CP-violating  phases
by one to two orders of magnitude. We find that the dependence of $d_D$ on the 
underlying QCD-sector CP-odd sources is closest to $d_{\rm Hg}$ and is complementary to $d_n$.
Moreover, in addition to the improvement in precision, $d_D$ has a significant advantage 
over $d_{\rm Hg}$ due to the rather transparent nuclear physics 
in the former and thus smaller theoretical uncertainties. 
Consequently, the experiment will be able to probe classes of supersymmetric models 
which escape the current EDM bounds.

We now proceed to analyze the deuteron EDM $d_D$, 
defined via the interaction of the deuteron spin $\vec{I}$ with 
an electric field, ${\cal H}=-d_D\, \vec{I}\cdot \vec{E}$, 
working upwards in energy scale. Starting at the nuclear level,
the deuteron EDM receives contributions from a singlet combination of 
the constituent proton and neutron EDMs, but also arises 
due to meson (predominantly pion) exchange between the nucleons
with CP-odd couplings at one of the meson-nucleon vertices. Thus, we
can represent the EDM as 
\begin{equation}
 d_D = (d_n + d_p) + d_D^{\pi NN},
\end{equation}
where the third term includes the meson-exchange contribution and
depends on the CP-odd pion nucleon couplings, 
\begin{equation}
 {\cal L}_{\rm CP \!\!\!\!\!\!\slash} \; = \bar{g}^{(0)}_{\pi NN} \bar N \ta^a N \pi^a 
 + \bar{g}^{(1)}_{\pi NN} \bar N N \pi^0.
\end{equation}
In a recent analysis, Khriplovich and Korkin \cite{kk} (see also \cite{ss})
showed that $d_D^{\pi NN}$ receives a dominant contribution from the 
isospin-triplet coupling $\bar{g}^{(1)}$. In a zero-radius
approximation for the deuteron wavefunction, the result
\begin{equation}
 d_D^{\pi NN} = -\frac{e g_{\pi NN} \bar g_{\pi NN}^{(1)}}{12\pi m_\pi} ~
                \frac{1+\xi}{(1+2\xi)^2}, 
\end{equation}
depends on the parameter $\xi=\sqrt{m_p \ep}/m_\pi$,
determined by the deuteron binding energy $\ep=2.23$ MeV. 
Numerically, this implies
\begin{equation}
 d_D^{\pi NN} \simeq  - (1.3 \pm 0.3) ~e~\bar g_{\pi NN}^{(1)}\, [{\rm GeV}^{-1}],
\end{equation}
a result that can be improved systematically, and the error correspondingly reduced \cite{kk}, 
with the use of more realistic deuteron wave functions. 

To make direct contact with models of CP violation, we require the 
dependence of $d_n$, $d_p$, and $\bar{g}^{(1)}$
on the parameters in the underlying CP-odd Lagrangian at 1 GeV. Up
to dimension five, the relevant hadronic operators are
the $\th$-term and  the EDMs and color EDMs (CEDMs) of quarks
\begin{equation}
 {\cal L}_{\rm CP \!\!\!\!\!\!\slash}\;
=\bar\th \frac{\al_s}{8\pi}G\tilde{G}
   -\frac{i}{2}\sum_{q=u,d,s}\left[ d_q \bar{q}F\si \ga_5 q 
+\tilde{d}_q \bar{q} g_s G\si \ga_5 q\right],
\end{equation}
where $G\tilde{G} \equiv \ep_{\mu\nu\rh\si}G^{\mu\nu a} G^{\rh\si a}/2$ and
$G\si \equiv t^a G^{\mu\nu a}\si_{\mu\nu}$. 
Note that the dimension-six Weinberg operator, $GG\tilde{G}$, as well as 
numerous four-quark operators, may, in certain models, also contribute at a 
similar level to the quark EDMs and CEDMs.

Models of new CP-violating physics can be cast into two main categories: (i) 
models that have no Peccei-Quinn (PQ) symmetry \cite{PQ} and exact CP or P symmetries at 
high energies and consequently $\bar\theta = 0$ at tree level; and (ii) models 
that invoke a Peccei-Quinn symmetry to remove any dependence of the 
observables on $\bar\theta$. 
In models of the first type, $\bar\theta$ generated by radiative corrections 
is likely to be the main source of EDMs.

To determine $d_D(\bar\theta)$, one may first try to make use of the chiral
techniques \cite{crewther} that determine the $\bar\theta$-induced 
pion-nucleon coupling constant, 
$\bar{g}^{(0)}_{\pi NN}(\bar\th) = m_* \bar\th f_\pi^{-1} 
\langle N|\bar{u} u - \bar{d} d|N\rangle$ (where $m_*=m_um_d/(m_u+m_d)$),
 and the one loop $O(m^2_\pi \log ~m_\pi)$ contribution to $d_n$. 
It is easy to see, however, that $d_D(\bar\theta)$ is incalculable 
within this approach because the chiral logarithms exactly cancel in 
the $d_n+d_p$ combination, and $\bar{g}^{(1)}(\bar\theta)=0$ unless isospin 
violating corrections are taken into account.

The cancellation between $d_n(\bar\th)$ and $d_p(\bar\th)$ does not hold in general.
To calculate $d_D(\bar\theta)$ we 
use leading order QCD sum-rule estimates which imply \cite{pr1},
\bea
 d_n(\bar\th) + d_p(\bar\th) &=&  \nonumber\\
   && \!\!\!\!\!\!\!\!\!\!\!\!\!\!\!\!\!\!\!\!\!\!\!\!\!\!\!\!\!
   - (2 \pm 0.8)\,\pi^2 \left(\frac{m_N}{\rm 1\, GeV}\right)^3 
 \frac{\langle \bar{q}q\rangle}{({\rm 1\, GeV})^3} m_* \ch e \bar\th,
   \label{dNtheta}
\eea
where 
$\langle \bar{q}\si_{\mu\nu} q \rangle_F = e_q\ch F_{\mu\nu} \langle \bar{q}q\rangle$
defines the magnetic susceptibility $\ch\sim -(6-9)$ GeV$^{-2}$ \cite{chi} of the 
vacuum, recently 
computed to be at the upper end of this range, $\ch=-N_c/(4\pi^2f_\pi^2)$,  
by Vainshtein \cite{vainshtein}. The subleading corrections to the sum rule
were computed and are of order 10-15\% \cite{pr1}, while the uncertainty in $\ch$
and freedom in the choice of nucleon interpolating current lead to a 
larger overall uncertainty of 30-40\% \cite{pr1}.

It turns out that despite an additional suppression factor,
the corresponding contribution to $\bar{g}^{(1)}(\bar\th)$ is not negligible and
contributes to $d_D$ at approximately the same level as (\ref{dNtheta}). 
To take it into account, we note 
that isospin violation arises  predominantly through $\et-\pi$ mixing as shown in Fig.~1(a). 
The inverted diagram of Fig.~1(b) provides 
at most a 10\% correction, due primarily to the small size of $g_{\et NN}$
and $\langle N| \bar{u}u - \bar{d}d | N\rangle$ relative to $g_{\pi NN}$ and 
$\langle N | \bar{u}u + \bar{d}d - 2 \bar{s}s | N \rangle$. 
Fig.~1(a) leads to the following result:
\begin{equation}
 \bar{g}_{\pi NN}^{(1)}(\bar\th) = \frac{m_* \bar\th}{f_\pi} 
   \frac{m_d - m_u}{4m_s} \langle N|\bar{u} u + \bar{d} d - 2\bar{s} s|N\rangle.
    \label{g1theta}
\end{equation}

\begin{figure}
\includegraphics[width=6.5cm]{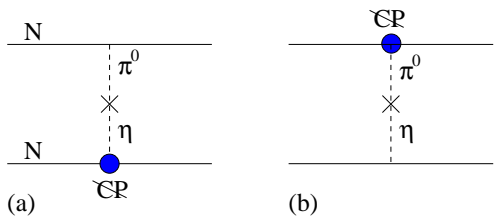}
\vspace*{0.3cm}
 \caption{\footnotesize Contributions to $d_D^{\pi NN}(\bar\th)$, with
  isospin violation through $\et-\pi$ mixing.  }
\label{f4} 
\end{figure}

Combining (\ref{dNtheta}) and (\ref{g1theta}), we obtain
\bea
 d_D(\bar\th) &=& -e \bar\th \left[
    2\pi^2 \frac{\ch m_*\langle \bar{q}q\rangle}{({\rm 1\, GeV})^3}
 \right.\nonumber\\
 && \!\!\!\!\!\!\!\!\!
 +\left.\frac{m_*}{m_s}\frac{(m_d-m_u)}{4f_\pi}
  \langle N|\bar{u} u + \bar{d} d - 2\bar{s} s|N\rangle
   \right], \label{dDtheta}
\eea
which numerically takes the form
\bea
 d_D(\bar\th) &\simeq& 
   -e \left[(3.5 \pm 1.4) + (1.4 \pm 0.4)\right] \nonumber \\
    && \;\;\;\;\;\;\;\;\;\;\;\times 10^{-3}\bar\th \; [{\rm GeV}^{-1}],
\label{dedmth}
\eea
using standard quark mass ratios \cite{leut}, 
and quark condensates over the nucleon (see e.g. \cite{bm}).
The second term in (\ref{dedmth}) arises from the CP-odd pion-nucleon interaction.

This result is interesting for several reasons. Firstly, if the projected experimental
sensitivity (\ref{bound}) is achieved, a null result for $d_D$ will imply 
\begin{equation}
 |\bar\th | < 3\times 10^{-11}, \label{thbnd}
\end{equation}
which represents an improvement of over an order of magnitude relative to
the best current bound arising from the limit on the neutron EDM. 
We note that the recent inclusion of many-body effects in the 
nuclear component of the calculation of $d_{\rm Hg}$ \cite{DS} has led to a significant 
{\em reduction} of $d_{\rm Hg}(\bar{g}^{(0)})$, thus relaxing the mercury EDM constraint 
on $\bar\theta$ by an order of magnitude. It is
also important to note that the two sources for $\bar\th$ in (\ref{dDtheta})
have quite different origins, and thus a cancellation would be unnatural.
Given the relatively good theoretical control over the contribution entering through
$\bar{g}^{(1)}$, the uncertainty in the estimate (\ref{dNtheta}) is of less concern.
The bound  (\ref{thbnd}) has important implications for 
solutions to the strong  CP problem within supersymmetry. 
In particular, the left-right symmetric SUSY models typically
predict $\bar\theta$ in the range  $10^{-8}-10^{-10}$ \cite{Babu:2001se},
allowing a direct probe via the $d_D$ experiment.

Introducing a PQ symmetry allows the
axion to relax to its minimum thereby rendering $\bar\th$ unobservable.
Adopting this approach, we are left with the
dimension five quark EDMs and CEDMs as the leading candidates for the position of
dominant CP-odd source. The
constituent EDMs of the proton and neutron receive contributions from
both of these operators, 
with the QCD sum-rules result 
(omitting for now the Weinberg operator) \cite{pr2}
\bea
 d_n(d_q,\tilde{d}_q)+d_p(d_q,\tilde{d}_q) &\simeq& (0.5 \pm 0.3) (d_u+d_d)\nonumber\\
   && \!\!\!\!\!\!\!\!\!\!\!\!\!\!\!\!\!\!\!\!\!\!\!\!\!\!\!\!\!\!\!\!\!\!\!\!\!\!
   \!\!\!\!\!\!\!\!\!\!\!\!\!\!\!\!\!\!\!\!\!
   -(0.6\pm 0.3)
  e \left[ (\tilde{d}_u-\tilde{d}_d) + 0.3 (\tilde{d}_u +\tilde{d}_d)\right],
     \label{dNedm}
\eea
where we have split the CEDM contribution into singlet and triplet combinations.
A possible contribution from $\tilde{d}_s$ is removed at this order under PQ
relaxation. The quoted errors have the same origin as those in (\ref{dNtheta})
for the dependence of $d_n$ and $d_p$ on $\bar{\th}$.

The triplet pion nucleon coupling $\bar{g}^{(1)}$ receives a dominant contribution
from the triplet combination $(\tilde{d}_u-\tilde{d}_d)$ of CEDMs, and
the ``best'' value for this coupling was recently determined using sum-rules
\cite{pospelov},
\begin{equation}
 \bar{g}^{(1)}_{\pi NN} \sim 2_{-1}^{+4}\times 10^{-12} \frac{\tilde{d}_u-\tilde{d}_d}
{10^{-26}~ {\rm cm}},
    \label{g1edm}
\end{equation}
with a rather large (overall) uncertainty due to an exact cancellation at the level
of vacuum factorization. We quote the non-Gaussian errors determined via parameter 
variation \cite{pospelov}.
Since this result enters without any additional isospin-violating suppression
factor, it numerically dominates the CEDM contribution to $d_D$. 
Combining (\ref{dNedm}) and (\ref{g1edm}), we find
\bea
 d_D(d_q,\tilde{d}_q) &\simeq& -e (\tilde{d}_u-\tilde{d}_d) 
  \left[5_{-3}^{+11} + (0.6 \pm 0.3)\right] \nonumber\\
    && \!\!\!\!\!\!\!\!\!\!\!\!\!\!\!\!\!\!\!\!\!\!\!\!\!\!\!\!\!\!\!\!\!\!\!\!
   - (0.2 \pm 0.1) e (\tilde{d}_u+\tilde{d}_d)+(0.5 \pm 0.3)(d_u+d_d),
\eea
where the constituent nucleon EDMs provide a 10\% correction to the
triplet CEDM contribution. We conclude from this result that
for models with $e\tilde d_i \sim d_i$ 
the deuteron EDM is predominantly sensitive to the 
triplet combination of CEDMs, as is the mercury EDM. 
Moreover, if the predicted precision is
achieved, its sensitivity to the triplet CEDM combination at the level of a
few$\times 10^{-28}$ $e\,$cm would represent an
improvement on the current mercury EDM bound by two orders of magnitude. 

We now turn to an analysis of the predicted sensitivity
to new CP-odd sources focusing on the minimal supersymmetric standard model
(MSSM) with universal boundary conditions at the GUT scale for all parameters 
except for those  in the Higgs sector. This exception allows us to satisfy
all phenomenological and cosmological constraints for a wide range of squark masses
while keeping the other parameters fixed \cite{nuhm}. In this case, there 
are two CP violating phases, identified 
with the phases of the $\mu$ parameter in the superpotential
 and the phase of a common trilinear soft-breaking
term $A_0$.

In Fig.~2, we plot the EDMs as a function of  
the left-handed down squark mass by varying $m_0$ from 0.25 - 10 TeV, while 
keeping $m_{1/2}$ (as well as the other input parameters)
fixed. For this choice of parameters, the light Higgs mass is about 120 GeV
and the lightest neutralino is a mixed gaugino/Higgsino state.  
The curves begin at $\tilde m_{d_L} \sim 1.2$ TeV corresponding to
$m_0 = 250$ GeV with $m_{1/2} = 600$ GeV.
In this figure, the theoretical average values of the neutron, thallium and 
mercury EDMs are normalized to their current experimental limits, while 
$d_D$ is normalized to $3\times 10^{-27}~e~{\rm cm}$. The theoretical error bands 
are generally very narrow on these log-scale plots and are not shown.
For low $\theta_{\mu(A)}$,  the EDMs scale with $\theta$ and therefore the
results for other (small) choices of $\theta_{\mu(A)}$ can be deduced from
this figure. We immediately see that the projected 
sensitivity of $d_D$ to squark masses extends beyond 10 TeV, 
well beyond that of the existing bounds or the reach of colliders
in the foreseeable future.
Note that the dips observable in the plot of $d_{\rm Hg}$ for $\th_\mu\neq 0$
are due primarily to cancellations between quark CEDM and electron EDM contributions.

\begin{figure}
\includegraphics[width=8cm]{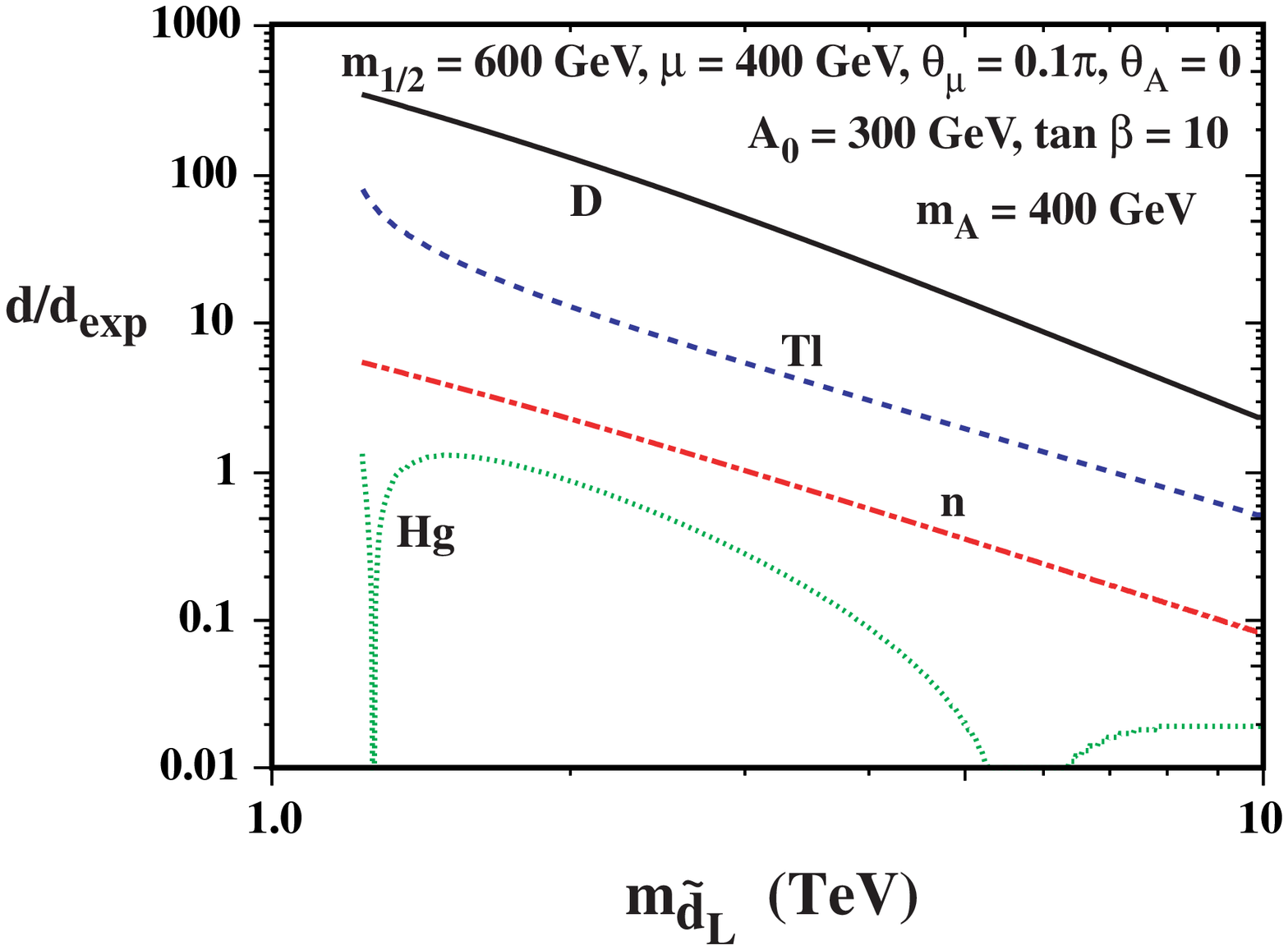}\\
\includegraphics[width=8cm]{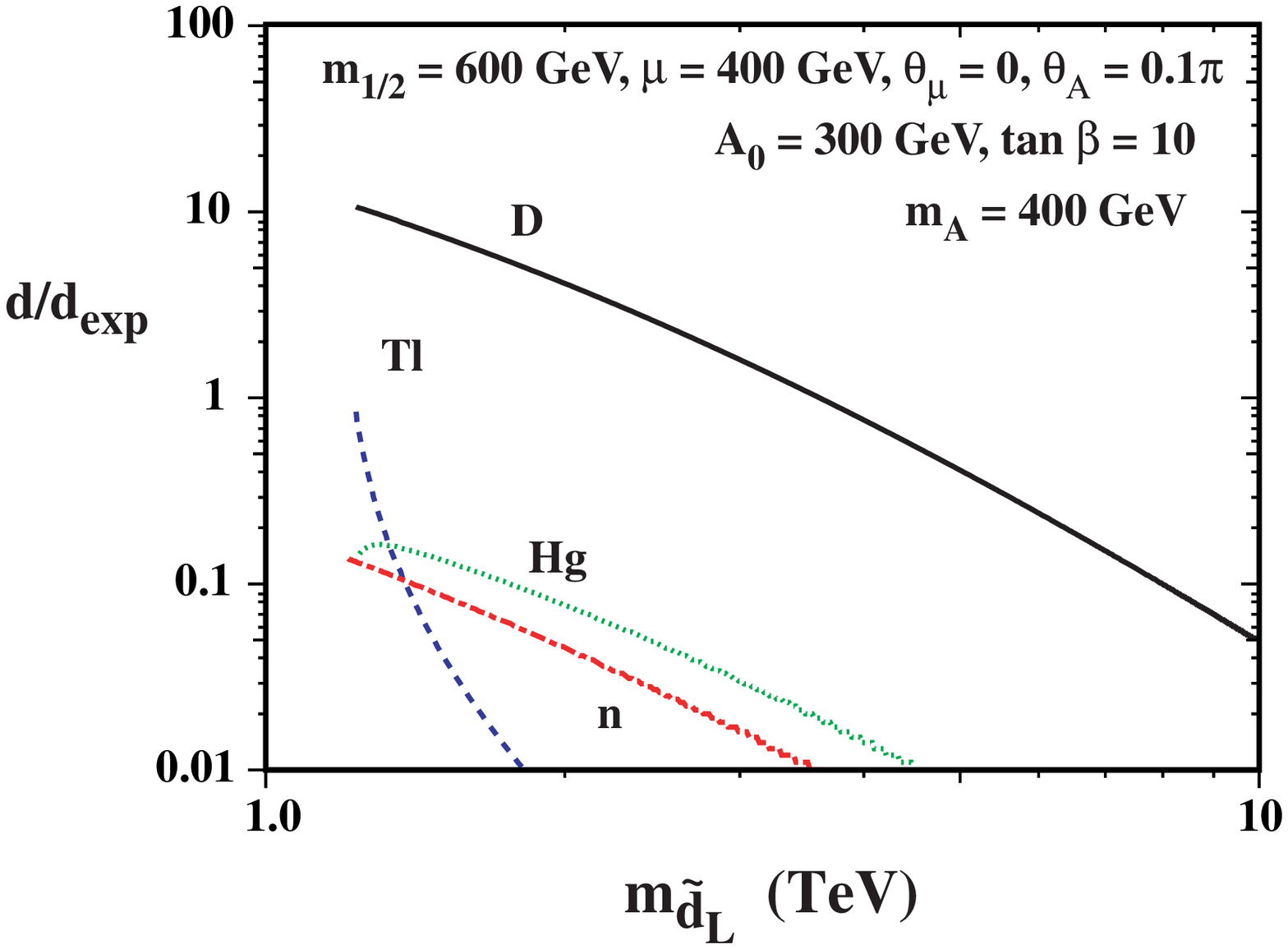}
\caption{\footnotesize The EDMs of the deuteron (black), mercury (green),
 the neutron (red), and thallium (blue) as a function of the SUSY soft breaking scalar 
 mass $m_0$, displayed in terms of the left-handed down squark mass. 
In a)  $\th_A=0, ~\theta_\mu=\pi/10$ and in b)
 $\th_A=\pi/10, ~\theta_\mu=0$. The EDM is normalized to the experimental constraint 
in each case.
} 
\label{edmthmu} 
\end{figure}

The $d_D$ experiment 
will also be able to probe a popular solution to
the SUSY CP problem, the ``decoupling'' scenario. 
This framework assumes that the sfermions of  
the first two generations have masses in the multi--TeV range 
thus suppressing the one--loop EDM contributions to an acceptable level and allowing
CP-odd phases to be of order one \cite{Nath:dn}. To satisfy the cosmological 
constraints on dark matter abundance \cite{fos},
and to avoid excessive fine-tuning in the Higgs sector, the masses
of the third generation sfermions should be near the electroweak scale. 
The Weinberg operator is then generated at two-loop order, providing 
the primary contribution to $d_D$ \cite{Weinberg,Demir:2002gg}:
\begin{equation}
d_D ~ \simeq ~ d_n(w) +d_p(w) ~ \sim ~  e\, w \times 20~{\rm MeV},
\end{equation}
where $w$ is the coefficient of the Weinberg operator evaluated at 1 GeV.
 The Weinberg operator provides a negligible contribution to $d_D^{\pi NN}$
due to additional chiral suppression and isospin violating factors in 
$\bar{g}^{(1)}_{\pi NN}(w)$.
Presently, order one CP-violating phases in this framework are barely compatible with the
experimental constraint on $d_n$ \cite{Abel:2001vy}. 
Therefore, an improvement in the experimental precision by a factor
of 10 or more, to the level of $10^{-27}$ $e\,$cm,  
would provide a crucial test for these models.
Failure to observe $d_D$ would necessarily imply that the CP-violating phases are small
contrary to the primary assumptions of the model.

\begin{figure}
\includegraphics[width=8cm]{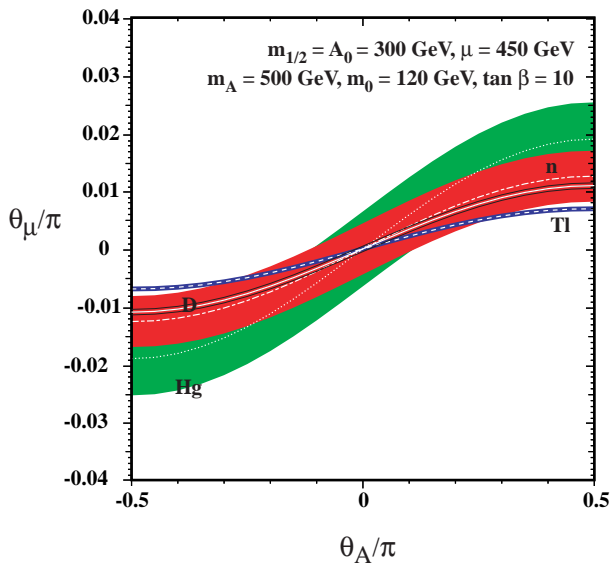}
 \caption{\footnotesize Bands of $|d|\leq d_{\rm exp}$ in the
 $\theta_A$ -- $\theta_\mu$
 plane for $A_0=m_{1/2}=300$ GeV, and $m_0=120$ GeV (with the same color-coding as in 
 Fig.~\ref{edmthmu}). The width of
 the deuteron band normalized to $3\times 10^{-27}~e~{\rm cm}$
 is too small to be visible on the plot and is artificially 
 widened by a factor of 10. 
}
\label{theta} 
\end{figure}

Next, we analyze constraints on the SUSY CP-violating phases  $\th_A$, $\th_\mu$
with the superpartner
mass scales fixed as shown in Fig. \ref{theta}.
This is a CMSSM point (the Higgs soft masses are unified with
other sfermion masses) with a relatively low Higgs mass of 114 GeV 
\cite{nuhm}.
We observe 
that $d_D$ combined with 
the thallium constraint 
can put tight bounds on both phases  including $\th_A$ that is otherwise poorly 
constrained.
An improvement of the bound on the triplet CEDM combination
by a factor of 30 or more would allow one to probe 
SUSY CP-odd phases of size $10^{-3}$ or below ($10^{-2}$ or so
for the $A$--terms). 
In a number of theoretically
motivated scenarios, phases of this size are naturally expected.
In particular, if the $A$--terms are hermitian at the GUT scale
as happens in the left--right and other models,
RG running  induces small phases in the diagonal elements.
For a variety of textures, the CEDMs of the light quarks are
of order $10^{-27}$ cm \cite{Abel:2000hn},
and thus observable at the upcoming experiment.

Finally, we consider the sensitivity of $d_D$ to the 
dimension 6 operators, $C_{ij}\bar q_i q_i  
\bar q_j i \gamma_5 q_j$, which may be important in 
two Higgs doublet models, left-right symmetric models, and certain  
supersymmetric scenarios. Typically, $C_{ij}$ can be parametrized as 
$C_{ij} = c Y^{\rm SM}_i Y^{\rm SM}_jM_h^{-2}, $
where $Y^{\rm SM}_{i(j)}$ are the SM quark Yukawa couplings, $M_h$ is the mass of the  
(lightest) Higgs boson, and the coefficient $c$ is model dependent.
Existing EDM bounds are sensitive to $C_{ij}$ only 
with the help of an enhancement at large $\tan\beta$, $c \sim \tan^2\beta$ 
or $\tan^3\beta$ \cite{ltb}, or in the 
top quark sector where $C_{tq}$ induces $w$ and/or light quark (C)EDMs via the
Barr-Zee mechanism \cite{BZ}. The projected sensitivity to $d_D$ 
would in contrast probe $C_{ij}$ for all quark flavors down 
to $c\sim 0.01-0.1$ for $M_h\sim 100$ GeV, thus providing  
valuable constraints even for $\tan \beta \sim O(1)$.

In conclusion, we have presented an analysis of the deuteron EDM
in terms of the relevant Wilson coefficients and 
studied  the implications of a
$d_D$ measurement at the level of a few$\times 10^{-27}~e\,$cm. 
We have shown that this would lead to a 
factor of 10 to 100 gain in sensitivity to various CP violating sources  
of dimension 4, 5 and 6. This has important consequences for supersymmetry
and other scenarios for physics  beyond the Standard Model.

{\bf Acknowledgments:}
We thank M. Voloshin for illuminating discussions. 
The work of KAO  was supported in part
by DOE grant DE--FG02--94ER--40823. 
A.R. thanks the Uni. of Minnesota and the Uni. of Victoria for hospitality 
while this work was in progress.

\end{document}